\documentclass[aps,nofootinbib]{revtex4}
\usepackage{amsmath,amssymb,amsbsy,amsfonts,latexsym,graphicx,hyperref}
\usepackage{color}
\usepackage{rotating}
\usepackage{graphicx,subfigure,color,array,slashed,wasysym}
\usepackage{caption}
\usepackage{dcolumn}
\usepackage{bm}
\usepackage[mathlines]{lineno}
\usepackage{hyperref}

\usepackage{bbold}

\newcommand{\be}{\begin{equation}}
\newcommand{\ee}{\end{equation}}
\newcommand{\ba}{\begin{array}}
\newcommand{\ea}{\end{array}}
\newcommand{\bea}{\begin{eqnarray}}
\newcommand{\eea}{\end{eqnarray}}

\newcommand{\dd}{\mathrm{d}}

\usepackage{multirow}

\begin{document}

\begin{center}
{\bf Doped Holographic Superconductors in Gubser-Rocha model}\\

\vspace{1.6cm}

Ziyi Zhao $^{a}$, Wenhe Cai $^{a,b,* }$\let\thefootnote\relax\footnotetext{* whcai@shu.edu.cn}, Shuta Ishigaki $^{a}$\\
\vspace{0.8cm}

$^a${\it Department of Physics, Shanghai University, Shanghai 200444,  China}\\
$^b${\it Shanghai Key Laboratory of High Temperature Superconductors, Department of Physics, 
Shanghai University, Shanghai 200444, China}
\vspace{1.6cm}

\begin{abstract}
\noindent ABSTRACT: We construct a doped  holographic superconductor in the Gubser-Rocha model, and realize a superconducting dome in the middle of the temperature-doping phase diagram. It is worth noting that unlike the previous researches, the profile of our dome shrinks inward near zero temperature. From the numerical observation for the coupling dependence of the phase diagram, we find that the coupling between the two gauge fields plays a crucial role in the formation of dome. We also analytically calculate the DC conductivity of the normal phase of the system in the momentum dissipation and obtain the resistivity which is proportional to temperature. The AC conductivity is calculated numerically.
\vspace{0.8cm}\\
\noindent keywords: AdS-CFT Correspondence, High-$T_c$ superconductor, AC/DC conductivities, Gubser-Rocha geometry
\end{abstract}

\end{center}

\maketitle
\newpage
\section{Introduction}
The thermodynamic and transport properties of non-Fermi liquids such as high temperature cuprate superconductors are strongly different from those described by the standard Fermi liquid theory\cite{carlson2002concepts}. There is still no satisfactory theoretical framework to describe them so far. The AdS/CFT correspondence, also known holographic duality\cite{Maldacena_1999,Gubser_1998,witten1998anti}, provides a useful tool for high-$T_c$ cuprate strange metals and other strongly correlated systems\cite{hooft2009dimensional,hartnoll2018holographic,zaanen2015holographic}. Recently, holographic duality has brought some new breakthroughs and discussions in condensed matter physics, such as: holographic superconductivity model\cite{Hartnoll_2008,Cai2015IntroductionTH,Ge_2010,10.1143/PTP.128.1211}, strange metal\cite{Hartnoll2010TowardsSM,Kim2012HolographicQC}, linear resistivity\cite{Jeong_2022,Jeong_2018}, Hall angle\cite{Pal2011ModelBI,Ahn2023InabilityOL}.
\par There are some similarities in the temperature-doped phase diagrams of many unconventional superconductors, such as high-temperature cuprates\cite{yuan2022scaling,Cai2015IntroductionTH,Hartnoll_2009}, with an antiferromagnetic phase, a superconducting phase, a metallic phase and a striped phase competing and coexisting with each other. How to construct a holographic theoretical model to reproduce and understand such a phase diagram becomes an important problem. In recent years, holographic two-current model has attracted attention\cite{Huang_2020,Kiritsis_2016,Baggioli_2016,Cai_2021,Rogatko_2018,bigazzi2012unbalanced,ZHANG2021136178,Seo_2017}. It has been pointed out that the model has a corresponding Mott insulator\cite{mott1936electrical,mott1936resistance,lee2006doping,ling2015building}, and there are two spintronic “up" and “down" currents which can be regarded as independent entities at low temperatures\cite{bigazzi2012unbalanced,Iqbal_2010,jin2011link}. These models include two gauge fields in bulk describing the two currents of the dual boundary field theory. Two independent conserved currents are associated with two different chemical potentials or charged densities, and their ratio defines the “doping" variable x, i.e. $\text{x}=\rho_A/\rho_B$. Recently, Kiritsis et al.\cite{Kiritsis_2016} added the related bulk fields dual to the operators of the boundary field theory to simulate the phase diagram that the normal phase, superconducting phase, antiferromagnetic phase and fringe phase compete with each other and coexist in the temperature-doped plane, in which the superconducting phase appears in the dome-shaped region in the middle of the phase plane. Ref.\cite{Baggioli_2016} introduces a neutral axion field of \cite{Baggioli_2015} to break the translational symmetry of the system, so that the normal phase has a finite DC conductivity distinguished from the superconducting phase. Their results also demonstrate that Kiritsis's superconducting dome still exists even with broken translational symmetry. Subsequent studies use this model to discuss the effects of quantum critical points on dome under hyperscaling violation geometry\cite{Cai_2021}. They reconsidered the role of the first class of charges with density $\rho_A$ in the temperature-doping phase diagram to form the Mott insulator, and the second class of charges with density $\rho_B$ to form additional charges by “doping".

\par On the other hand, previous studies construct holographic models with hyperscaling violating factor\cite{zhang2022excited,ge2017hyperscaling}. Particularly, Gubser-Rocha model is characterized by the
dynamical critical exponent $z \rightarrow \infty$ and hyperscaling violating exponent $\theta \rightarrow -\infty$. It implies the entropy is proportional to temperature near zero temperature with local quantum criticality\cite{Gubser2010PeculiarPO,balm2022tlinear}. It is similar to more realistic strange metals\cite{Ren_2023,Dom_nech_2010}. In order to simulate a more realistic holographic superconductor, in this paper, we construct a holographic superconducting dome in Gubser-Rocha model with two charges. One kind of charge is non-movable, which contribute to the half filled state of Mott insulator. One kind of charge is movable, which contribute to the state near the Fermi surface. The ratio of these two charges is the doping parameter. When the ratio increases, there is a dome-shaped superconducting region which is experimentally observed. That is, the critical temperature rises first and then decreases with the increase of doping. In order to make conductivity realistic, we also introduce holographic superconductor with momentum dissipation by breaking translational symmetry\cite{Baggioli_2015,Liu_2022,Liu_2023}. By calculating the DC conductivity of our model, we prove that the normal phase of the model satisfies the linear-$T$ resistivity relation\cite{Jeong_2022} in the presence of momentum dissipation. We also numerically calculate the AC conductivity\cite{fu2023charge,kim2015gauge,Konoplya:2009hv} of the normal phase with momentum dissipation. We also calculate spin conductivity $\gamma$ of our model, and as a result we observe behavior similar to the conductivity of Mott insulators. We get that there is indeed a superconducting dome phase in the middle of the $\text{x}$-$T$ plane in the Gubser-Rocha model, and unlike previous results, its profile shrinks inward at near zero temperature.
\par This paper is organized as follows. In section 2, we set up the doped holographic superconductor, based on the Gubser-Rocha model with broken translational symmetry. In section 3, we obtain the linear-$T$ resistivity analytically and AC conductivities numerically in the normal phase in order to explore the two-current model. In section 4, the normal phase of the bulk system was observed to become unstable at $T_c$, then developing a nontrivial profile of the scalar field $\chi$, which means a superconducting phase transition. We study superconducting instability with the critical temperature $T_c$ by numerically solving the motion equation of $\chi$ in the normal phase background, and draw the superconducting dome phase in temperature-doping plane. We conclude with discussions at section 5.

\section{Construction of The doped superconductor in Gubser-Rocha model}
Inspired by ref.\cite{Baggioli_2016}, we try to construct a holographic superconducting dome in Gubser-Rocha model\cite{Gubser2010PeculiarPO}. We consider the following action \eqref{eq1}:
\begin{equation}\label{eq1}
 \begin{aligned}
    S &=\int \sqrt{-g}\left[\mathcal{R}+\frac{6}{L^2} \cosh \varphi-\frac{1}{4}\left({\mathrm{e}} ^{\varphi}+Z_A(\chi)\right) A_{\mu \nu}A^{\mu\nu}-\frac{1}{4}\left({\mathrm{e}} ^{\varphi}+Z_B(\chi)\right) B_{\mu \nu}B^{\mu\nu}-\frac{1}{2} Z_{A B}(\chi) A_{\mu \nu} B^{\mu \nu}\right.\\
    &\left.-\frac{3}{2}\left(\partial_\mu \varphi\right)^2-H(\chi)\left(\partial_\mu \theta-q_A A_\mu-q_B B_\mu\right)^2-V_{i n t}(\chi)-\Sigma_{I=x, y}\left(\partial \phi_I\right)^2-\frac{1}{2}\left(\partial_\mu \chi\right)^2\right] {\mathrm{d}} ^4 x,
    \end{aligned}
\end{equation}
where we set the gravitational constant $16 \pi G=1$. The scalar field $\varphi$ is the ‘dilaton' that makes the entropy  proportional to temperature near zero temperature. The $A_{\mu \nu}$ and $B_{\mu \nu}$ stand for the field strengths of the gauge fields $A_{\mu}$ and $B_{\mu}$ which provide the finite chemical potentials. We introduce a perturbative charge complex scalar $\psi 
=\chi \mathrm{e} ^{i \theta}$ for superconducting instability, $\chi$ and $\theta$ are the amplitude
and phase of the charged complex scalar field, respectively. $Z_A$, $Z_B$ and $Z_{AB}$ are nontrivial
coupling, and $V_{int}$ is potential term for
the scalar field. We will fix these terms
later. The two axions $\phi_I$: $\phi_x = mx$, $\phi_y = my$ break the translational invariance. 
\par The corresponding equations of motion with the action \eqref{eq1} are obtained, as follows. The scalar field $\chi$ 's equation is
\begin{equation}\label{eq32}
\begin{aligned}
    &\nabla_\mu \nabla^\mu \chi-\partial_\chi V_{i n t}(\chi)\\
	&-\frac{1}{4} \partial_\chi Z_A(\chi) A^2-\frac{1}{4} \partial_\chi Z_B(\chi) B^2-\frac{1}{2} \partial_\chi Z_{A B}(\chi) A_{\mu \nu} B^{\mu \nu}-\partial_\chi H(\chi)\left(\partial_\mu \theta-q_A A_\mu-q_B B_\mu\right)^2=0  .
\end{aligned}
\end{equation}
The equations of motion of the other matter fields are
\begin{equation}
    \label{eq33}
3 \nabla_\mu \nabla^\mu \varphi +\frac{6}{L^2} \sinh \varphi-\frac{1}{4} {\mathrm{e}} ^{\varphi} A^2-\frac{1}{4} {\mathrm{e}} ^{\varphi} B^2=0 ,
\end{equation}
\begin{equation}\label{eq34}
    \nabla_{\mu} \left[\left({\mathrm{e}} ^{\varphi}+Z_A\right) A^{\mu \nu}\right]+\nabla_{\mu}\left[\left({\mathrm{e}} ^{\varphi}+Z_{A B}\right) B^{\mu \nu}\right]
    +2 q_A H(\chi)\left(\partial^\nu \theta-q_A A^\nu-q_B B^\nu\right)=0,
\end{equation}
\begin{equation}\label{eq35}
    \nabla_{\mu} \left[\left({\mathrm{e}} ^{\varphi}+Z_B\right) B^{\mu \nu}\right]+
    \nabla_\mu\left[\left({\mathrm{e}} ^{\varphi}+Z_{A B}\right) A^{\mu \nu}\right] \\
    +2 q_B H(\chi)\left(\partial^\nu \theta-q_A A^\nu-q_B B^\nu\right)=0,
\end{equation}
and $\nabla_{\mu}\nabla^{\mu} \phi_I = 0$ for the axions. The Einstein’s equation is given by
\begin{equation}\label{eq31}
\begin{aligned}
    &\mathcal{R}_{\mu \nu}-\frac{1}{2}\left({\mathrm{e}} ^{\varphi}+Z_A\right)  A_\mu^{~\beta} A_{\nu \beta}-\frac{1}{2}\left({\mathrm{e}}^{\varphi}+Z_B\right)  B_\mu^{~\beta} B_{\nu \beta}-\frac{1}{2} Z_{A B}\left(g^{\alpha \beta} A_{\mu \beta} B_{\nu \alpha}+g^{\alpha \beta} B_{\mu \beta} A_{\nu \alpha}\right) \\
    &-\frac{1}{2} \partial_\mu \chi \partial_\nu \chi-H(\chi)\left(\partial_\mu \theta-q_A A_\mu-q_B B_\mu\right)\left(\partial_\nu \theta-q_A A_\nu-q_B B_\nu\right)-\frac{3}{2} \partial_\mu \varphi \partial_\nu \varphi-\sum_{I=x, y} \partial_\mu \phi_I \partial_\nu \phi_I \\
    &-\frac{1}{2} g_{\mu \nu}\left[\mathcal{R}+\frac{6}{L^2} \cosh \varphi-\frac{1}{4}\left({\mathrm{e}}^{\varphi}+Z_A(\chi)\right) A_{\mu \nu} A^{\mu\nu}-\frac{1}{4}\left({\mathrm{e}}^{\varphi}+Z_B(\chi)\right) B_{\mu \nu} B^{\mu\nu}-\frac{1}{2} Z_{A B}(\chi) A_{\mu \nu} B^{\mu \nu}\right. \\
    &\left.-\frac{3}{2}\left(\partial_\mu \varphi\right)^2-H(\chi)\left(\partial_\mu \theta-q_A A_\mu-q_B B_\mu\right)^2-V_{i n t}(\chi)-\Sigma_{I=x, y}\left(\partial \phi_I\right)^2-\frac{1}{2}\left(\partial_\mu \chi\right)^2\right]=0 .
    \end{aligned}
\end{equation}
In the following part, the couplings and potential are considered as:
\begin{equation}\label{eq29}
    Z_A(\chi)= \frac{a \chi^2}{2},~~Z_B(\chi)=\frac{b \chi^2}{2},~~Z_{AB}(\chi)= \frac{c \chi^2}{2 },~~H(\chi)=\frac{n \chi^2}{2},~~V_{int}(\chi)= \frac{M^2 \chi^2}{2 }.
\end{equation}
And we define the $U(1)_{A,B}$ charges to be:
\begin{equation}
    \label{eq30}
    q_A=1,~~~~~~q_B=0.
\end{equation}
Although this value is very specific, Kiritsis et al. have verified that this is feasible\cite{Kiritsis_2016}. 
\par To obtain the solutions for this holographic system \eqref{eq1}, we assume the metric ansatz
\begin{equation}\label{eqd}
\begin{aligned}
 {\mathrm{d}} s^2 &=-f(r) {\mathrm{d}} t^2+\frac{{\mathrm{d}} r^2}{f(r)}+g(r)\left({\mathrm{d}} x^2+{\mathrm{d}} y^2\right) , \\
    A_t&=A_t(r),~~~B_t=B_t(r),\\
    \phi_x&=mx,~~~~\phi_y=my,\\
    \varphi &=\varphi(r),~~~~\chi=\chi(r),~~~~\theta\equiv 0,  \\
    \end{aligned}
\end{equation}
where $x, y$ are spatial coordinates on the boundary and $r$ denotes the radial bulk coordinate. The boundary is located at $r\rightarrow \infty$, the horizon is located at $r=r_0$. The Hawking temperature of the black brane is given by
\begin{equation}\label{eqe}
    T=\left.\frac{f^{\prime}(r)}{4 \pi}\right|_{r_0}.
\end{equation}
\par In normal phase ($\chi\equiv 0$), our model admits the following solution,
\begin{equation}\label{eq2}
\begin{aligned}
    \varphi(r) &=\frac{1}{2} \ln (1+Q/r), ~~g(r)=r^{1/2}\left(r+Q\right)^{3/2},~~ \\
    f(r) &=r^{1/2}\left(r+Q\right)^{3/2} (1-\frac{(\mu^2+\mu_B^2)\left(Q+r_0\right)^2}{3 Q (r+Q)^3} -\frac{m^2}{(r+Q)^2}) .\\
    A_t(r)&=\mu(1 -\frac{r_0+Q}{r+Q}),~~~B_t(r)= \mu_B(1 -\frac{r_0+Q}{r+Q}),\\
    \phi_x&=mx,~~~ \phi_y=my,\\
    \end{aligned}
\end{equation}
where $r_0$ denotes the horizon radius, $m$ denotes a strength of the momentum relaxation, $Q$ is a physical parameter. The density of the charge carriers is denoted by $\rho _A=\mu(Q+r_0)$, which is dual to the gauge field
$A_\mu$, while the density of doped charge $\rho_B=\mu_B(Q+r_0)$ is dual to the gauge field $B_\mu$. The doping ratio is given by:
\begin{equation}\label{eq9}
    \text{x} = \frac{\rho_B}{\rho_A}=\frac{\mu_B}{\mu}.
\end{equation}
The chemical potentials $\mu$ and $\mu_B$ satisfy the following equations,
\begin{equation}\label{eq3}
\mu=\sqrt{\frac{3 Q (Q+r_0)}{1+\text{x}^2} \left(1-\frac{m^2}{(Q+r_0)^2}\right)},
\end{equation}
\begin{equation}\label{eqb3}
\mu_B=\text{x}\sqrt{\frac{3 Q (Q+r_0)}{1+\text{x}^2} \left(1-\frac{m^2}{(Q+r_0)^2}\right)}.
\end{equation}
The Hawking temperature is given by
\begin{equation}\label{eq4}
    T=r_0^{\frac{1}{2}} \frac{\left(3\left(Q+r_0\right)^2-m^2\right)}{4 \pi\left(Q+r_0\right)^{\frac{3}{2}}}.
\end{equation}

\section{Conductivities in The Normal Phase}
\par We consider the momentum dissipation and calculate AC/DC conductivities of our model in the normal phase. Because of the symmetry of the x-y plane, without loss of generality, we only consider the disturbance in the $x$ direction. The system includes two $U(1)$ fields $A$ and $B$ as whose dual operators in the boundary theory denoted as $J_A$ and $J_B$ describing two currents, respectively. We also write corresponding external electric fields and
conductivities as $E_A$, $E_B$ and $\sigma_A$, $\sigma_B$,
respectively. In general, the external field $E_A$ also contributes to the current $J_B$. This process is reciprocal. The associated conductivity is called as $\gamma$. $\sigma_A$ and $\sigma_B$ are interpreted as electric conductivity and spin-spin conductivity, and relatively $\gamma$ is spin conductivity\cite{ZHANG2021136178}. The heat current $\mathcal{Q}$ is also coupled with these
two currents. They satisfy Ohm's law, which is expressed as
\begin{equation}\label{eq22}
\left(\begin{array}{c}
J_A \\
\mathcal{Q} \\
J_B
\end{array}\right)=\left(\begin{array}{ccc}
\sigma_A & \alpha T & \gamma \\
\alpha T & \kappa T & \beta T \\
\gamma & \beta T & \sigma_B
\end{array}\right)\left(\begin{array}{c}
E_A \\
-\nabla T / T \\
E_B
\end{array}\right) ,
\end{equation}
where $\kappa$ is the thermal
conductivity. $\alpha$ and $\beta$ are
called as thermo-electric and thermospin conductivities, respectively. This non-diagonal matrix is symmetric as a result of the time-reversal symmetry. We compute the conductivities $\sigma_A$, $\gamma$ and $\sigma_B$ in the normal phase of this system. We will discuss the details in the following subsections.

\subsection{DC conductivities}
Using the methods in \cite{Chen_2017}, we obtain analytic expressions for the several DC conductivities in the normal phase: 
\begin{equation}\label{eq5}
\sigma_A=\left.{\mathrm{e}}^{\varphi}+\frac{r^{1/2}(r+Q)^{3/2}A_t'^2(r)}{2m^2}{\mathrm{e}}^{2 \varphi}\right|_{r=r_0},
\end{equation}
\begin{equation}\label{eq6}
\sigma_B=\left.{\mathrm{e}}^{\varphi}+\frac{r^{1/2}(r+Q)^{3/2}B_t'^2(r)}{2m^2}{\mathrm{e}}^{2 \varphi}\right|_{r=r_0},
\end{equation}
\begin{equation}\label{eq7}
\gamma=\left.\frac{r^{1/2}(r+Q)^{3/2}A_t'(r)B_t'(r)}{2m^2}{\mathrm{e}}^{2 \varphi}\right|_{r=r_0}.
\end{equation}
For convinience, we definite scaled variables by
\begin{equation}\label{eq8}
\tilde{Q}\text{:=}\frac{Q}{r_0}, \quad \tilde{m}\text{:=}\frac{m}{r_0}, \quad \tilde{T}\text{:=}\frac{T}{r_0}, \quad \tilde{\mu}\text{:=}\frac{\mu}{r_0}, \quad
\tilde{\mu_B}\text{:=}\frac{\mu_B}{r_0}.
\end{equation}
We also define two physical dimensionless quantities\cite{Jeong_2022}:
\begin{equation}\label{eq10}
\bar{T}\text{:=}\frac{\tilde{T}}{\tilde{\mu}}=\frac{3(1+\tilde{Q})^2-\tilde{m}^2}{4\sqrt{3}\pi \sqrt{\frac{\tilde{Q}(1+\tilde{Q})^2} {1+\text{x}^2}((1+\tilde{Q})^2-\tilde{m}^2) }  } ,
\end{equation}
\begin{equation}\label{eq11}
\bar{m}\text{:=}\frac{\tilde{m}}{\tilde{\mu}}=\sqrt{\frac{(1+\tilde{Q}) \tilde{m}^2}{\frac{3 \tilde{Q}}{1+\text{x}^2}\left((1+\tilde{Q})^2-\tilde{m}^2\right)}}.
\end{equation}
Using \eqref{eq10},\eqref{eq11}, one can express $\tilde{Q}$ as a function of $\bar{T}$ and $\bar{m}$, i.e., $\tilde{Q}(\bar{T},\bar{m})$. The DC conductivities \eqref{eq5}-\eqref{eq7} are expressed in terms of $\tilde{Q}$, $\bar{m}$ and x as
\begin{equation}\label{eq12}
\sigma_A=\sqrt{1+\tilde{Q}}+\frac{\sqrt{1+\tilde{Q}}}{2\bar{m}^2},
\end{equation}
\begin{equation}
    \label{eq12b}
\sigma_B=\sqrt{1+\tilde{Q}}+\text{x}^2\frac{\sqrt{1+\tilde{Q}}}{2\bar{m}^2},
\end{equation}
\begin{equation}
    \label{eq12c}
    \gamma=\text{x} \frac{\sqrt{1+\tilde{Q}}}{2\bar{m}^2}.
\end{equation}
The conductivity $\sigma_A$ and $\sigma_B$ can be divided into the following two cases by the values of $\bar{T}$ and $\bar{m}$.\\
\par If $\bar{T}~\ll1~\text{for~given}~\bar{m}$, 
\begin{equation}\label{case1}
\tilde{Q} \sim \frac{3\left(1+2\bar{m}^2\right)^2}{16 \pi^2\left(1+3 \bar{m}^2\right) \bar{T}^2} 
\end{equation}
\begin{equation}\label{eq13}
 \sigma_{\mathrm{A}} \sim \frac{\sqrt{3}\left(1+2\bar{m}^2\right)^2}{8 \pi \bar{m}^2 \sqrt{1+3 \bar{m}^2}} \frac{1}{\bar{T}},~~~~~~
 \sigma_{\mathrm{B}} \sim \frac{\sqrt{3}\left(1+2\bar{m}^2\right) (\text{x}^2+2 \bar{m}^2)}{8 \pi \bar{m}^2 \sqrt{1+3 \bar{m}^2}} \frac{1}{\bar{T}}.
\end{equation}
\par If $\bar{m}~\gg1~\text{for~given}~\bar{T}$, 
\begin{equation}\label{case2}
    \tilde{Q} \sim \frac{\bar{m}^2}{4 \pi^2 \bar{T}^2},
\end{equation}
\begin{equation}\label{dcb}
\sigma_{\mathrm{A}} \sim \frac{\bar{m}}{2  \pi }\frac{1}{\bar{T}},~~~~~~
\sigma_{\mathrm{B}} \sim \frac{\bar{m}}{2  \pi }\frac{1}{\bar{T}}.
\end{equation}

As can be seen from the above equation, the resistivity ($\rho=1/\sigma_{DC}$) of this model is proportional to temperature in some limits.

\subsection{AC conductivities}

For calculating AC conductivities numerically, we turn on the bulk fluctuations for $A_x$, $B_x$ and $g_{tx}$ around the background \eqref{eq2}, which serve as the sources for the currents $J_x^A$ and $J_x^B$, as well as the stress energy tensor component $T_{tx}$ in the dual boundary field theory.
For our purpose, we assume that the fluctuations depend only on $t$ and $r$ coordinates.
We set $g_{rx}=0$ by a gauge choice. 
In a case with the momentum dissipations,
the axions are also coupled with those fluctuations. By virtue of the symmetry, we
consider only the fluctuation for $\phi_x$. We consider the ansatz of the fluctuations as
\begin{equation}\label{eq15}
\delta A_x(t, r)=\int_{-\infty}^{\infty} \frac{\mathrm{d} \omega}{2 \pi} \mathrm{e}^{-i \omega t} a_x(\omega, r),
\end{equation}
\begin{equation}\label{eq16}
\delta B_x(t, r)=\int_{-\infty}^{\infty} \frac{\mathrm{d} \omega}{2 \pi} \mathrm{e}^{-i \omega t} b_x(\omega, r),
\end{equation}
\begin{equation}\label{eqa}
\delta \phi_x(t,r)=\int_{-\infty}^{\infty} \frac{\mathrm{d} \omega}{2 \pi} \mathrm{e}^{-i \omega t} \Phi_x(\omega, r),
\end{equation}
\begin{equation}\label{eq17}
\delta g_{t x}(t, r)=\int_{-\infty}^{\infty} \frac{\mathrm{d} \omega}{2 \pi} \mathrm{e}^{-i \omega t} \frac{r^2}{r_0^2} h_{t x}(\omega, r).
\end{equation}

We obtain five equations by linearizing the full equations of motion (see Appendix A for details) but only four are independent due to the gauge symmetry in the system.
Near the horizon, the solutions are expanded as
\begin{equation}\label{eq23}
\begin{aligned}
& a_x=(r-r_0)^{\lambda}\left(a_x^{(I)}+a_x^{(II)}(r-r_0)+\cdots\right), \\
& b_x=(r-r_0)^{\lambda}\left(b_x^{(I)}+b_x^{(II)}(r-r_0)+\cdots\right), \\
& \Phi_x=(r-r_0)^{\lambda}\left(\Phi_x^{(I)}+\Phi_x^{(II)}(r-r_0)+\cdots\right),\\
& h_{tx}=(r-r_0)^{\lambda+1}\left(h_{t x}^{(I)}+h_{t x}^{(II)}(r-r_0)+\cdots\right),
\end{aligned}
\end{equation}
where
$\lambda = - i\omega/(4\pi T)$
and it corresponds to the incoming-wave boundary condition.
By solving the equations of motion near the horizon, we obtain one constraint on $h_{tx}^{(I)}, a_x^{(I)}, b_{x}^{(I)}$ and $\Phi_{x}^{(I)}$, which reduces a degree of freedom.
As a result, the number of independent incoming-wave solutions are only $3$ nevertheless there are $4$ fields.
The reason of this mismatch of the degree of freedom is because of the residual gauge freedom of the diffeomorphism invariance, as we will explain soon later.
Near the boundary ($r \rightarrow \infty$), the asymptotic behaviors of the fluctuations are
\begin{equation}\label{eq24}
\begin{aligned}
    a_x =& a_x^{(0)}+\frac{1}{r} a_x^{(1)}+\cdots,&
    b_x =& b_x^{(0)}+\frac{1}{r} b_x^{(1)}+\cdots,\\
    \Phi_x =& \Phi_x^{(0)}+\frac{1}{r} \Phi_x^{(1)}+\cdots,&
    h_{tx} =& h_{tx}^{(0)}+\frac{1}{r^3} h_{tx}^{(1)}+\cdots.
\end{aligned}
\end{equation}
According to the AdS/CFT dictionary, the leading terms ($a_x^{(0)}$, $b_x^{(0)}$, $\Phi_x^{(0)}$, $h_{tx}^{(0)}$) correspond to the sources and the subleading terms ($a_x^{(1)}$, $b_x^{(1)}$, $\Phi_x^{(1)}$, $h_{tx}^{(1)}$) are considered as the responses.
However, due to the mismatch of the degrees of freedom, we cannot impose the Dirichlet conditions for the $4$ source values ($a_x^{(0)}$, $b_x^{(0)}$, $\Phi_x^{(0)}$, $h_{tx}^{(0)}$) at the boundary with the incoming-wave conditions at the same time.
To cure this problem, there are two methods are known, as follows.%
\footnote{
    One can maybe obtain master equations for master fields in a similar way to Sec.~2.7 of \cite{Hartnoll_2009} for the RN-AdS$_4$, or \cite{Wu:2017exh} with the momentum dissipation.
    However, we do not try to find such a combination of the fields here.
}

The first method is imposing appropriate constraints on the boundary values from the gauge symmetry in the boundary theory.\cite{Donos:2013eha}
We refer this method as boundary constraint method.
The boundary values from both of the background and the fluctuations can be written as
\begin{equation}
    \Psi_{\text{bou}}
    :=
    \lim_{r\to\infty}
    \begin{pmatrix}
        A + \delta A\\ B + \delta B\\
        \phi_{I} + \delta\phi_{I}\\
        \frac{1}{r^2} \left(
            g_{ij} + \delta g_{ij}
        \right)
    \end{pmatrix}
    =
    \begin{pmatrix}
        \mu_A \dd{t}\\ \mu_B \dd{t}\\ m x_I\\ \eta_{ij}
    \end{pmatrix}
    +
    \int\frac{\dd{\omega}}{2\pi}\mathrm{e}^{-i\omega t}
    \begin{pmatrix}
       a^{(0)}\\ b^{(0)}\\ \Phi_I^{(0)}\\
        r_0^{-2} h_{ij}^{(0)}
    \end{pmatrix},
\end{equation}
where $i,j = t,x,y$ denotes the indices of the boundary coordinates.
As we have mentioned, we cannot impose the Dirichlet conditions for all of the boundary values as long as imposing the incoming-wave condition at the horizon.
Instead of these values, we can regard the different values as physical sources of the fluctuations, obtained by a coordinate change in the boundary theory.
Considering an infinitesimal coordinate transformation generated by $\xi$ which has the same order to the fluctuations, we obtain
\begin{equation}
    \Psi_{\text{bou}}
    \to
    \hat{\Psi}_{\text{bou}}
    =
    \left(1 + \mathcal{L}^{\eta}_{\xi}\right)
    \begin{pmatrix}
        \mu_A \dd{t}\\ \mu_B \dd{t}\\ m x_I\\ \eta_{ij}
    \end{pmatrix}
    +
    \int\frac{\dd{\omega}}{2\pi}\mathrm{e}^{-i\omega t}
    \begin{pmatrix}
       a^{(0)}\\ b^{(0)}\\ \Phi_I^{(0)}\\
        r_0^{-2} h_{ij}^{(0)}
    \end{pmatrix}
    =
    \begin{pmatrix}
        \mu_{A}\dd{t}\\ \mu_{B}\dd{t}\\ m x_I\\ \eta_{ij}
    \end{pmatrix}
    +
    \int\frac{\dd{\omega}}{2\pi}\mathrm{e}^{-i\omega t}
    \begin{pmatrix}
        \hat{a}^{(0)}\\ \hat{b}^{(0)}\\ \hat{\Phi}_I^{(0)}\\ r_0^{-2} \hat{h}_{ij}^{(0)}
    \end{pmatrix},
\end{equation}
where $\mathcal{L}^{\eta}_{\xi}$ denotes the Lie derivative associated with $\eta_{ij}$.
We have neglected the Lie derivatives of the fluctuations because they are quadratic order of the fluctuations and $\xi$.
($\hat{a}^{(0)}$, $\cdots$, $\hat{h}^{(0)}_{ij}$) denote the boundary values of the fluctuations in the new frame.
Now, we consider $\xi^{i} = (0,\tilde{\zeta}(t),0)$ which depends on $t$ only.
The nonzero components of the Lie derivatives are
\begin{equation}\label{eq:Lie_deriv}
    \mathcal{L}_{\xi}^{\eta} \phi_x = \xi^{i}\partial_{i}\phi_x = m \tilde{\zeta},\quad
	(\mathcal{L}_{\xi}^{\eta} \eta_{ij})_{tx}
    =
    (\partial_{i} \xi_{j} + \partial_{j} \xi_{i})_{tx}
	= \partial_{t} \tilde{\zeta}.
\end{equation}
Using these, we obtain the relation
\begin{equation}
    \Phi_x^{(0)} + m \zeta = \hat{\Phi}_x^{(0)},\quad
    h_{tx}^{(0)} - i\omega r_0^2 \zeta = \hat{h}_{tx}^{(0)},
\end{equation}
where $\zeta$ is the Fourier coefficient of $\tilde{\zeta}$.
Imposing $\hat{\Phi}_x^{(0)} = 0$, we need to fix $\zeta = - \Phi_x^{(0)}/m$, and obtain the constraint on the boundary values
\begin{equation}\label{eq:boundary_constraint}
    h_{tx}^{(0)} + \frac{i\omega}{m} r_0^2 \Phi_{x}^{(0)} = \hat{h}_{tx}^{(0)}.
\end{equation}
To compute the electric or spin conductivities, we also impose $\hat{h}_{tx}^{(0)} = 0$.
With this constraint, the number of the degrees of freedom becomes $3$ in the UV, and it agrees with those in the IR.
Remark that $h_{tx}^{(0)}=\Phi_x^{(0)}=0$ is not imposed, here.

The second method is using an extra solution along the residual gauge orbit.\cite{kim2014coherent,kim2015gauge}
We call this method as residual gauge symmetry (RGS) method.
The similar method was also utilized in \cite{Amado:2009ts}.
The diffeomorphism generated by $\xi^{\mu}=(0,\hat{\zeta}(t),0,0)$ acts even in the whole bulk geometry.
The nonzero components of the Lie derivatives are
\begin{equation}\label{eq:bulk_Lie_deriv}
	\mathcal{L}_{\xi} \phi_{x} = m \tilde{\zeta},\quad
	(\mathcal{L}_{\xi} g_{\mu\nu})_{tx}
	= g(r) \partial_{t} \tilde{\zeta},
\end{equation}
where $\mathcal{L}_{\xi}$ denotes the Lie derivative associated with $g_{\mu\nu}$. 
Our gauge $g_{rx}=0$ still holds under this transformation, so it is called RGS.
One can see that Eq.~(\ref{eq:bulk_Lie_deriv}) reduces to Eq.~(\ref{eq:Lie_deriv}) in the $r\to\infty$ limit.
The transformation can generates an extra solution from the trivial zero solution, i.e,
\begin{equation}\label{eq:extra_solution}
	a_x = 0,\quad
	b_x = 0,\quad
	\Phi_x= m \zeta_0,\quad
	h_{tx} = -i \omega \frac{r_0^2}{r^2} g(r) \zeta_0,
\end{equation}
where $\zeta_0$ is a normalization constant.
This solution does not satisfy the incoming-wave condition at the horizon but it is acceptable because it corresponds to the unphysical degree of freedom.
Gathering this solution with the incoming-wave solutions, we recover $4$ degrees of freedom.
Then, we can impose the Dirichlet conditions for each $4$ boundary values independently.

According to the AdS/CFT dictionary, the AC conductivities are given by the following formulas \cite{bigazzi2012unbalanced}
\begin{equation}\label{eq26}
\begin{gathered}
\sigma_A
=-\left.\frac{i}{\omega} \frac{a_x^{(1)}}{a_x^{(0)}}
\right|_{b_x^{(0)}=0},\quad
\sigma_B
=-\left.\frac{i}{\omega} \frac{b_x^{(1)}}{b_x^{(0)}}
\right|_{a_x^{(0)}=0},\\
\gamma
=-\left.\frac{i}{\omega} \frac{a_x^{(1)}}{b_x^{(0)}}
\right|_{a_x^{(0)}=0}
=-\left.\frac{i}{\omega} \frac{b_x^{(1)}}{a_x^{(0)}}
\right|_{b_x^{(0)}=0}.
\end{gathered}
\end{equation}
The boundary conditions for $h_{tx}$ and $\Phi_x$ at $r\to\infty$ must be chosen appropriately depending on the method curing the degrees of freedom.
For the first method, the constraint (\ref{eq:boundary_constraint}) should be imposed.
For the second, RGS method, the Dirichlet conditions $h_{tx}^{(0)} = \Phi_x^{(0)} =0$ should be imposed.
In the both, the suitable solution can be obtained by taking the linear combination of the basis solutions.
See Appendix \ref{sec:numerical_details} for more details.
We have checked both method always give same results for the electric/spin conductivities.


In the following, we divide the calculation of AC conductivities into two cases depending on whether momentum dissipation exists or not. We compare them and study the effect on conductivity after breaking translational symmetry.

\subsubsection{A case without momentum dissipation}
We study the AC conductivity in a case without momentum dissipation, $m=0$, i.e., the system preserves the translational invariance. We consider the symmetey between conductivities in normal phase. 
As reported in a different two-currents model \cite{bigazzi2012unbalanced}, our system also enjoys the symmetry about $\text{x}$:
\begin{equation}\label{eq27}
\sigma_A(\text{x},\omega)=\sigma_B(\frac{1}{\text{x}},\omega),~~~~~
\gamma(\text{x},\omega)=\gamma(\frac{1}{\text{x}},\omega)
\end{equation}
We have checked the above symmetry numerically. We show the results compared with the case where there is momentum dissipation later.

\par
\begin{figure}[htpb]
\begin{centering}
\subfigure
{\includegraphics[scale=0.75]{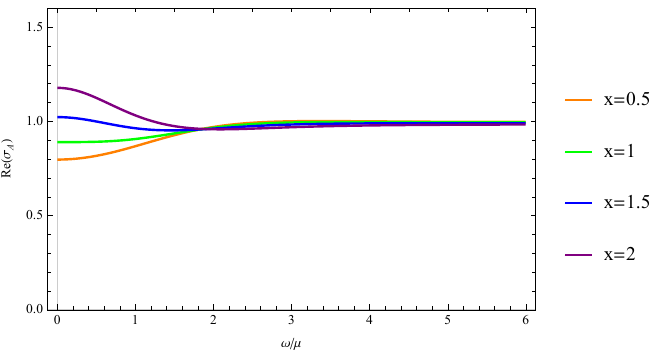}}
\subfigure
{\includegraphics[scale=0.75]{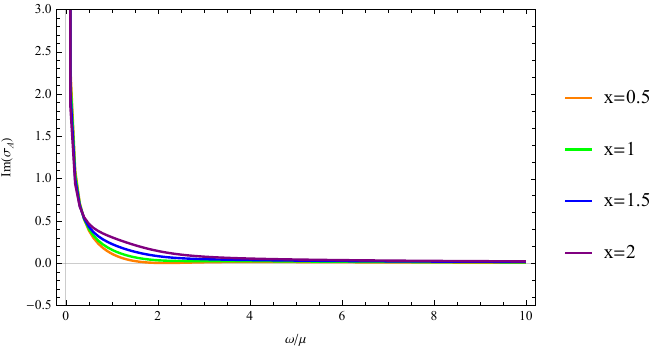}}
\par\end{centering}
\caption{\label{fig:Figure3} The real part(left) and the imaginary part(right) of the conductivity $\sigma_A$ for various $\text{x}=0.5, 1, 1.5, 2$ (orange, green, blue, purple) without momentum dissipation}. The temperature is fixed at $T/\mu=0.31$. 
\end{figure}

 We compare the effects of different doping parameters x on the conductivity. Figure \ref{fig:Figure3} shows the AC conductivity for various $\text{x}$ in a case without momentum dissipation.
In Fig.\ref{fig:Figure3}, we find that with the increase of doping $\text{x}$, the asymptotic value of real part of the conductivity at $\omega=0$ also increases due to the increases of the density of carriers.
Remark that the DC conductivity is actually infinite due to the presence of the Dirac delta at $\omega=0$ in this case.
According to \cite{davison2015incoherent}, such a finite part of the real part of the conductivity in the vicinity of $\omega=0$ can be understood as the incoherent conductivity $\sigma_Q$. The analytic expression for $\sigma_Q$ will be obtained by performing the low-frequency expansion but we leave it as a future study. For larger $\text{x}$, one can see the broad peak around $\omega = 0$ in the real part of the $\sigma_A$. It implies there is a Drude-like peak with a finite width in addition to the delta peak of the translational symmetry.

\subsubsection{A case with momentum dissipation}
Next, we consider the case where momentum dissipation is introduced, i.e. $m\neq 0$. Although the translational symmetry has been broken due to the presence of the axions, our results show the conductivities $\sigma_A$, $\sigma_B$ and $\gamma$ in the normal phase still satisfy the symmetry \eqref{eq27}, see Appendix C for details.

\par
\begin{figure}[htpb]
\begin{centering}
\subfigure[ Re($\sigma_A$) vs. $\omega/\mu$ at $m/\mu = 1$.]{
			\label{fig:Figure7a}
{\includegraphics[scale=0.75]{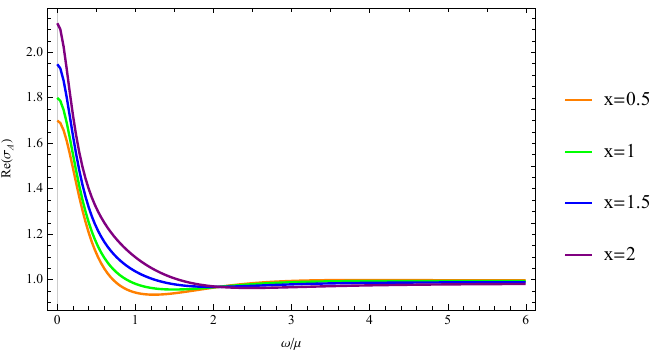}}}
\subfigure[ Im($\sigma_A$) vs. $\omega/\mu$ at $m/\mu = 1$.]{
			\label{fig:Figure7b}
{\includegraphics[scale=0.75]{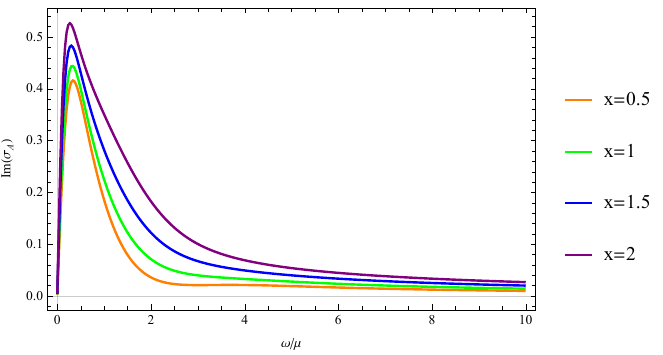}}}
\par\end{centering}
\caption{\label{fig:Figure7} The conductivities $\sigma_A$ change with the different doping parameter $\text{x}=0.5, 1, 1.5, 2$ (orange, green, blue, purple) at $m/\mu=1$. The temperature is fixed at $T/\mu=0.31$.}
\end{figure}

 We fix $m/\mu=1$ and compare the effect of increasing x on the conductivity $\sigma_A$ in Fig.\ref{fig:Figure7}. Due to the presence of the momentum dissipation, the Drude peak is broaden and the DC conductivity becomes the finite value given by \eqref{eq5}. We have checked that the DC limit of the numerics at $\omega=0$ agrees with the analytic value $\sigma_A$ in \eqref{eq5}. 

\par
\begin{figure}[htpb]
\begin{centering}
\subfigure[ Re($\gamma$) without momentum dissipation]{
			\label{fig:Figure10a}
{\includegraphics[scale=0.75]{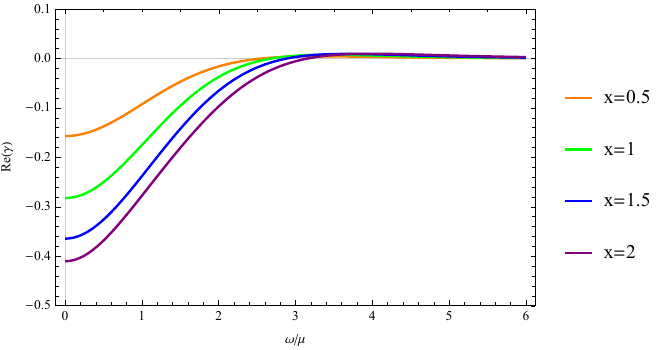}}}
\subfigure[ Re($\gamma$) with momentum dissipation $m/\mu=1$]{
			\label{fig:Figure10b}
{\includegraphics[scale=0.75]{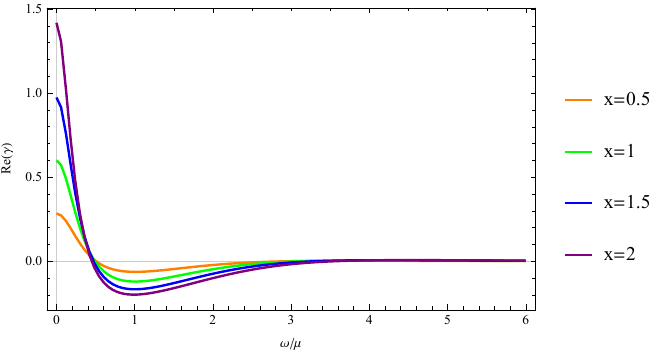}}}
\par\end{centering}
\caption{\label{fig:Figure10} The conductivities $\gamma$ change with the different doping parameter $\text{x}=0.5, 1, 1.5, 2$ (orange, green, blue, purple) at $m/\mu=1$. The temperature is fixed at $T/\mu=0.31$.}
\end{figure}

 We compare the conductivity $\gamma$ with different x in different cases in Fig.\ref{fig:Figure10}. Our results show that, the values of positive conductivities $\gamma$ at $\omega=0$ increase with the increase of x inthe case of $m/\mu=1$. On the contrary, the values of negative conductivities decrease accordingly without momentum dissipation. This observation is very interesting, and we consider that it is the result of competition between the momentum dissipation that breaks the translational symmetry and the chemical potential.

\begin{figure}[htpb]
\begin{centering}
\subfigure[ Re($\sigma_A$). There is a delta function at $\omega=0$ for $m/\mu=0$ but not shown.]{
			\label{fig:Figure6a}
{\includegraphics[scale=0.7]{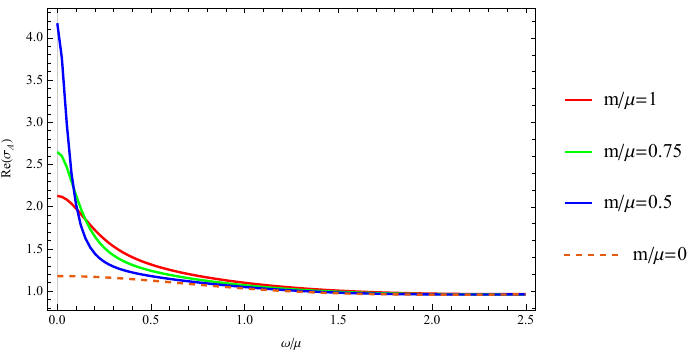}}}
\subfigure[ Im($\sigma_A$). There is a $1/\omega$ pole for $m/\mu = 0$ corresponding to a delta function in (a).]{
			\label{fig:Figure6b}
{\includegraphics[scale=0.7]{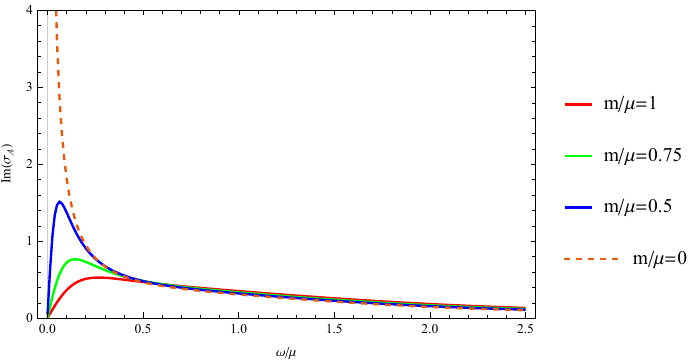}}}
\par\end{centering}
\caption{\label{fig:Figure6} The conductivities $\sigma_A$ change with the different dissipation intensity $m/\mu=1, 0.75, 0.5, 0$ (red, green, blue, dashed) at $T/\mu=0.31$.}
\end{figure}

Figure \ref{fig:Figure6} shows how the conductivity $\sigma_A$ changes with different dissipation intensity $m/\mu$ at $\mathrm{x}=2$. Figure \ref{fig:Figure6a} is the real part and Fig.\ref{fig:Figure6b} is the imaginary part of the conductivity $\sigma_A$. We find that as $m/\mu$ increases, the value of $\mathrm{Re}(\sigma_A$) at $\omega=0$ is smaller and the Drude peak disappears.

\par
\begin{figure}[htpb]
\begin{centering}
{\includegraphics[scale=0.75]{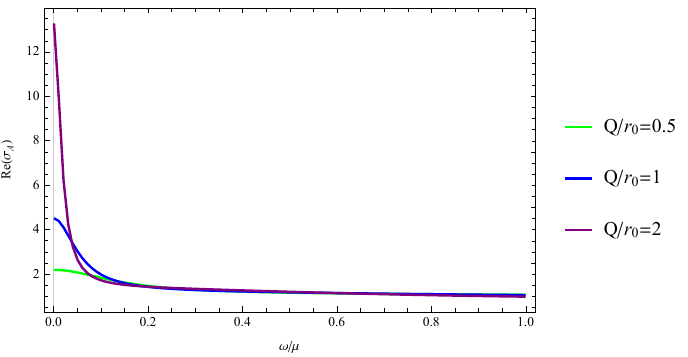}}
\par\end{centering}
\caption{\label{fig:Figure12} The real parts of conductivities $\sigma_A$ change with $\omega/\mu$ at different parameter $Q/r_0=0.5, 1, 2$ (green, blue, purple) corresponding to $T/\mu = 0.445, 0.312, 0.219$, respectively. We set $m/r_0 = 0.5$.}
\end{figure}

 We present the the real part of the AC conductivities $\sigma_A$ as functions of $\omega/\mu$ for various $Q/r_0$, in order to see influence of the parameter $Q$ related to the dilaton profile.
We fix $\text{x} = 2$ and $m/r_0=0.5$, here.
As $Q/r_0$ increases, the conductivity decreases at low frequencies and tends to a stable value at high frequencies.
The AC conductivity is given by a constant value when $Q=0$ corresponding the constant dilaton profile.
A case $Q=0$ corresponds to the constant dilaton profile, in which the AC conductivity is given by a constant value.
The result in Fig.\ref{fig:Figure12} agrees with the expected behavior.
Note that $T$ and $\mu$ functions of $r_0, Q, x$ and $m$, so they varies when we change $Q$.
For fixed $m$ and $x$, different $Q/r_0$ corresponds to different $T/\mu$.

\section{The Superconducting Dome}
In this section, we investigate the phase diagram in $\mathrm{x}$-$T$ plane in our model. At the critical temperature, the spontaneous breaking of $U(1)$ symmetry is dual to the condensation of scalar field $\chi$ in the bulk, and the superconducting instability corresponds to developing a non-trivial scalar $\chi$. In order to investigate whether the boundary system exhibits a superconducting phase, we study the instability of the scalar hair around the normal
phase of the dual bulk system. We solve the linearized equation of motion about $\chi$ in the background of \eqref{eq2} to determine the superconducting phase of the boundary system.
Here, we assume that the phase transition is second order for simplicity, so we regard the onset of the charged scalar instability as phase transition point.
\footnote{
More precisely, we have to study the nonlinear condensation to see whether the phase transition is the second order.
However, holographic models without nonlinear potential term exhibit the second order phase transition usually.
If the transition is a first order, the instability edge we investigate here does not agree with the phase transition points, but rather the edge of the metastable region of the normal phase.
}
\par The phase transition is related to the formation of scalar hair around the normal phase.
When the temperature is below the critical temperature $T_c$, the system becomes unstable and the scalar hair begins to develop. In the vicinity of the temperature at which the system developes non-trivial scalar hair of $\chi$, the value of $\chi$ should be small so we can consider it as a perturbation. Then we solve the linear motion equation \eqref{eq32} of the scalar field $\chi$ in the normal phase background without taking backreaction into consideration. By making a coordinate transformation of the metric form \eqref{eq2} i.e., $u = 1/r$, \eqref{eq32} yields
\begin{equation}\label{eq36}
    \chi ''+\left(\frac{f'}{f}+\frac{g'}{g}-\frac{2}{u}\right) \chi ' +\frac{\chi  \left(f\left(a u^2 A_t'^2+b u^2 B_t'^2+2 c u^2 A_t' B_t'-\frac{2 M^2}{u^2}\right)+2 n (q_A A_t+q_B B_t)^2\right)}{2f^2}=0.
\end{equation}
where ‘prime' stands for $\partial_u$.
\par To solve the equation \eqref{eq36}, we need to impose appropriate boundary conditions. In the IR limit, near the horizon, we impose the regular condition for the scalar field, it can be expressed as
\begin{equation}\label{eq37}
\chi(u)=\chi_h^{(I)}+\chi_h^{(II)}(u-u_h)+\frac{\chi_h^{(III)}}{2!}(u-u_h)^2+\cdots .
\end{equation}
In the UV, near $u =0 $, $\chi$ has an asymptotic expansion of
\begin{equation}\label{eq38}
    \chi(u)=\chi^{(-)}u^{3-\Delta}+\chi^{(+)}u^{\Delta}+\cdots ,
\end{equation}
where $\Delta$ is a larger root of $M^2=\Delta (\Delta-3)$, and we fix the scaling dimension $\Delta=5/2$ following \cite{Kiritsis_2016}. We also impose a condition to vanish the nonnormalizable term. As a result, the solution of this two-points boundary values problem is a static zero mode which indicates onset of the instability.
\par In the following, we fix the parameters \eqref{eq7} in the action. We also fix $\mu$ or $\rho_A$ as a scale depending on whether
considering in the grand canonical ensemble or the canonical ensemble, respectively. Here, we can also fix $Q = 1$ by virtue of the scaling symmetry without loss of generality.
In other words, we can regard quantities scaled by $Q$, such as $m/Q, u_h Q, T/Q, \mu/Q$, as new quantities. Then, the leaving parameters are the translational symmetry breaking parameter $m$, the doping x and the location of the horizon $u_h$. These are related to the temperature $T$ and the chemical potential $\mu$ by \eqref{eq4} and \eqref{eq3}, respectively. We look for the critical values $(\text{x},T)$ at which the boundary condition is satisfied for fixed $m$: the source coefficient 
$\chi^{(-)}$ of the field $\chi$ near the boundary expansion disappears there. We can find the multiple sets of $(\text{x},T)$, but the outermost set in the $\text{x}$-$T$ plane is expected to be dominant for the instability. We have also checked that the solution $\chi(u)$ for such a $(\text{x},T)$ has no node in its u-coordinate profile. In the following, we will only show the outermost set of $(\text{x},T)$ as a critical values.

\par
\begin{figure}[htpb]
\begin{centering}
{\includegraphics[scale=0.8]{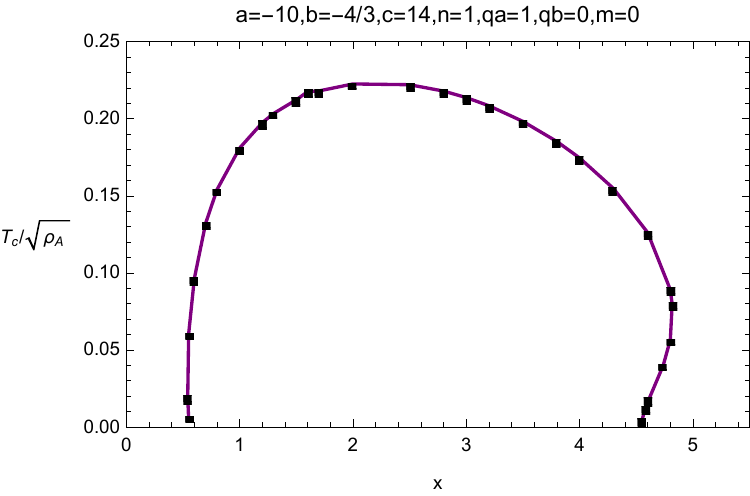}}
\par\end{centering}
\caption{\label{fig:Figure11} The critical temperature $T_c/\sqrt{\rho_A}$ vs. the doping parameter $\text{x}$ of the model \eqref{eq39} without momentum dissapation. The dots come directly from our numeric calculation. }
\end{figure}

\begin{figure}[htpb]
\begin{centering}
{\includegraphics[scale=0.8]{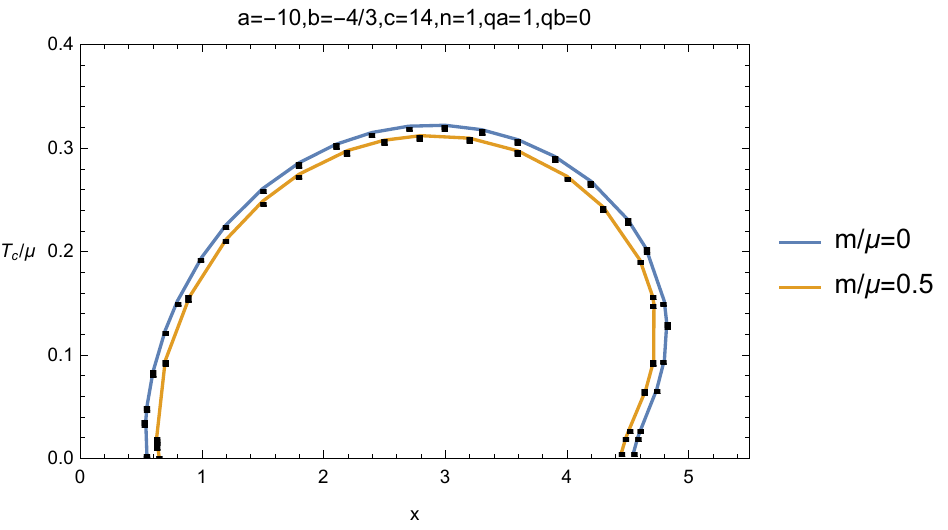}}
\par\end{centering}
\caption{\label{fig:Figure4} The phase diagram of the (x, $T_c/\mu$) plane of the model \eqref{eq39} at finite temperature when the translational symmetry breaking parameter is $m/\mu=0, 0.5$ (blue, orange). The dots come directly from our numeric calculation. }
\end{figure}

 Now, we test a set of model parameters with the following values:
\begin{equation}
    \label{eq39}
    a=-10,~b=-\frac{4}{3},~c=14,~n=1,~M^2=-\frac{5}{4}.
\end{equation}
Note that we have already fixed $q_A=1$ and $q_B=0$.
Figure \ref{fig:Figure11} shows the critical temperature $T_c$ as a function of $\text{x}$ for $m=0$, by settings $\rho_A$ as a scale, i.e., in the canonical ensemble.
In the grand canonical ensemble, we plot the critical temperature as a function of the doping parameter x by setting $\mu$ as a scale, as shown in Fig.\ref{fig:Figure4}. Here, we draw the phase diagram of the $\mathrm{x}$-$T$ plane with translational symmetry breaking parameter $m/\mu=0, 0.5$. In both cases, we find that dome shape does appear in the x-$T$ plane. It can indeed be seen that the critical temperature $T_c$ obtained by numerical calculation first increases and then decreases with the increase of doping, which is in line with the phase diagram characteristics of high temperature superconductors in reality. In the region near zero temperature, however, the phenomenon that the black dots near the endpoints shrink inward appears different from the previous model. As $r_0 \rightarrow 0$, i.e., $u_h \rightarrow \infty$, the Gubser-Rocha model can reach zero temperature. The black dots near the two endpoints whose $u_h$ are large in our model, which fit the zero temperature conditions of the original Gubser-Rocha model\cite{Gubser2010PeculiarPO}. Compared with former gravitational background researches, the Gubser-Rocha holographic high temperature-doped superconductor shows a difference in the region near zero temperature. We also prove that the introduction of translational symmetry breaking does not destroy the original superconducting dome structure in our extended model.

\par
\begin{figure}[htpb]
\begin{centering}
{\includegraphics[scale=0.75]{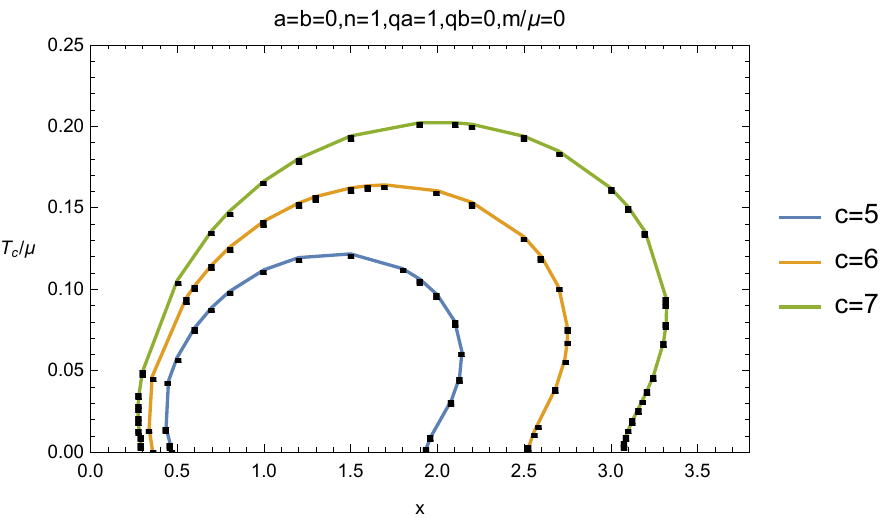}}
\par\end{centering}
\caption{\label{fig:Figure8} The phase diagram of the (x, $T_c/\mu$) plane. We fix $a = b = 0$, $m/\mu = 0$ and choose different $c$. The higher the value of $c$, the larger dome. }
\end{figure}

 We also discuss the influence of the value of coupling “$c$" on dome and obtain the same results as those obtained in \cite{Cai_2021}. We fix $a = b = 0$, $m/\mu = 0$ and choose different $c$. The result is shown in Fig.\ref{fig:Figure8}. Our result indicates that the higher the value of $c$, the larger the dome. The action brought by the coupling of the two gauge fields is obviously the decisive factor leading to the emergence of dome. 
 
\par With breaking translational symmetry, we reproduce the high-$T_c$ superconducting dome in a black brane whose entropy is proportional to temperature near zero temperature. Similarly to Fig.\ref{fig:Figure4}, in the vicinity of absolute zero temperature, a clear trend emerges as the inward shrinking of black dots near the endpoints. We also find that even if the coupling to a single gauge field is zero, relying on the cross term coupling, we can still obtain a superconducting dome. It is the main result of our paper. 

\section{Conclusion and discussion}
In this paper, we studied the holographic theory of building a dome region in the temperature-doping plane like a realistic high-$T_c$ superconductor. We attempt to realize a superconducting dome on the $\text{x}-T$ plane phase diagram in the holographic superconductors model, the Gubser-Rocha model. We first calculate the DC conductivities in the normal phase analytically. Our results show the resistivity is proportional to temperature with the momentum dissipation in normal phase. We then calculate conductivities associated to two gauge fields of our model in the normal phase numerically. We find there is a symmetry between $\sigma_A$, $\gamma$ and $\sigma_B$ with specific doping parameter x in normal phase. Our results indicate momentum dissipation can not break this symmetry.

\par We furthermore investigate the phase diagram for the normal and superconducting phases. One of the main progress of this paper is to show that the superconducting dome-shaped region exists in the extended Gubser-Rocha model in finite temperature with broken translational symmetry. This is a generalization of the previous work\cite{Cai_2021}. In comparison with other previous studies, the dome shrinks inward at both endpoints near zero temperature. This behavior might be caused by the presence of the dilaton field in our model. The parameter $Q$ is closely related to the profile of the dilaton, while it is also involved in the temperature and chemical potential, and the number of parameter is same as the Reissner-Nordstr\"om black hole case. Previous works show that the right boundary of dome gently tends outward to zero temperature, and the experimental data is between our results and the previous studies. Our work provides a new possibility for better simulation of the experimental phenomenon, which is worthy of further study. Moreover, our results show that larger coupling $c$ makes the dome higher in this background, which also verified the role of the three point interaction $-c \chi^2 A_{\mu \nu} B^{\mu \nu}$ under the hyperscaling violation geometry.

In this study, we employed the holographic model based on the Gubser-Rocha model for the purpose of constructing the holographic model of strange metals with a doping parameter.
As we have shown, the model can reproduce the typical behavior of the linear-$T$ resistivity in some limits, not involving magnetic field. However, the authors of \cite{Ahn2023InabilityOL} argued that the Gubser-Rocha model with linear axions can not capture all the strange metallic behaviors. We have to consider further improvements of the model to overcome the difficulties for reproducing the strange metallic behaviors by using the holography. In any case, the doping should be introduced for constructing realistic holographic superconducting model since it is a typical parameter in the unconventional superconductors.

\par We have constructed a holographic s-wave doped superconductor closer to the realistic background in a Gubser-Rocha black brane whose entropy is proportional to temperature near zero temperature. Yet most unconventional superconductors have d-wave symmetry, we hope to extend the present work to d-wave holographic superconductor in the future. It would be interesting to follow the lines that we are studying here.

\section*{Acknowledgements}
We would like to thank Xian-Hui Ge, Sang-Jin Sin, Blaise Gout\'eraux and Li Li for valuable comments and discussions. This work is supported by NSFC China (Grants No.\,12275166, No.\,11875184, No.\,12147158 and No.\,11805117) and NFSC-NFR joint program 12311540141.

\appendix

\section{The linearised equations in momentum space}
In this section, we write the linearised equations of motion for the fluctuations studied in section 3.
These are obtained as
\begin{equation}\label{eq18}
\frac{\omega^2 a_x(r)}{f(r)^2}+\frac{a_x^{\prime}(r) f^{\prime}(r)}{f(r)}+a_x^{\prime}(r) \varphi^{\prime}(r)+\frac{r^2 h_{t x}^{\prime}(r) A_t^{\prime}(r)}{r_0^2 f(r)}+a_x^{\prime \prime}(r)+h_{tx}(r)\left(\frac{2 r A_t^{\prime}(r)}{r_0^2 f(r)}+\frac{r^2 \varphi^{\prime}(r) A_t^{\prime}(r)}{r_0^2 f(r)}+\frac{r^2 A_t^{\prime \prime}(r)}{r_0^2 f(r)}\right)=0,
\end{equation}
\begin{equation}\label{eq19}
\frac{\omega^2 b_x(r)}{f(r)^2}+\frac{b_x^{\prime}(r) f^{\prime}(r)}{f(r)}+b_x^{\prime}(r) \varphi^{\prime}(r)+\frac{r^2 h_{t x}^{\prime}(r) A_t^{\prime}(r)}{r_0^2 f(r)}+b_x^{\prime \prime}(r)+h_{tx}(r)\left(\frac{2 r B_t^{\prime}(r)}{r_0^2 f(r)}+\frac{r^2 \varphi^{\prime}(r) B_t^{\prime}(r)}{r_0^2 f(r)}+\frac{r^2 B_t^{\prime \prime}(r)}{r_0^2 f(r)}\right)=0,
\end{equation}
\begin{equation}\label{eqb}
\Phi_x''(r)+(\frac{f'(r)}{f(r)}+\frac{g'(r)}{g(r)}) \Phi_x'(r)+\frac{\omega^2}{f(r)^2} \Phi_x(r)-\frac{i m r^2 \omega h_{tx}(r)}{r_0^2 f(r)^2 g(r)}=0,
\end{equation}
\begin{equation}\label{eq20}
\ h_{tx}(r)\left(\frac{2}{r}-\frac{g^{\prime}(r)}{g(r)}\right)+h_{tx}^{\prime}(r)+\frac{{\mathrm{e}}^{\varphi(r)} r_0^2}{r^2}\left(a_x(r) A_t^{\prime}(r)+b_x(r) B_t^{\prime}(r)\right) +\frac{2 i m r_0^2 f(r) \Phi_x^{\prime}(r)}{r^2 \omega} =0,
\end{equation}
\begin{equation}\label{eq21}
\begin{aligned}
h_{tx}^{\prime \prime}(r)+\frac{4 h_{tx}^{\prime}(r)}{r}+&\frac{{\mathrm{e}}^{\varphi(r)} r_0^2 }{r^2}(a_ x^{\prime}(r) A_t^{\prime}(r)+b_ x^{\prime}(r) B_t^{\prime}(r))-\frac{2 i m r_0^2 \omega }{r^2 f\mathrm(r)}\Phi_x(r)\\
+h_{tx}(r)
\Bigg(&\frac{2}{r^2}+\frac{3 }{f(r)}({\mathrm{e}}^{-\varphi(r)}+{\mathrm{e}}^{\varphi(r)})-\frac{f^{\prime}(r) g^{\prime}(r)}{f(r) g(r)}+\frac{g^{\prime}(r)^2}{2 g(r)^2}-\frac{3}{2} \varphi^{\prime}(r)^2\\
&+\frac{{\mathrm{e}}^{\varphi(r)}}{2 f(r)} (A_t^{\prime}(r)^2+B_t^{\prime}(r)^2)-\frac{f^{\prime \prime}(r)}{f(r)}-\frac{2 g^{\prime \prime}(r)}{g(r)}-\frac{2 m^2}{f(r) g(r)}\Bigg)=0.
\end{aligned}
\end{equation}
The last equation can be obtained from the others.
Thus, the set of the independent equations are given by \eqref{eq18}, \eqref{eq19}, \eqref{eqb} and \eqref{eq20}.

\section{Numerical details for the computation of the conductivities}\label{sec:numerical_details}
In this section, we provide some details for the numerical computation of the electric/spin conductivities.
The procedure is mostly same as those explained in \cite{Amado:2009ts,Kaminski:2009dh}.
However, we need to find the suitable solutions depending on the methods for curing the problem of the degrees of freedom.
Now, we write the bulk fluctuations as 4 components vector $\delta\Psi = (a_x, b_x, \Phi_x, h_{tx})^{T}$.
Imposing the incoming-wave condition at the horizon, we obtain $3$ independent solution ($\delta\Psi^{\{1\}}$,$\delta\Psi^{\{2\}}$,$\delta\Psi^{\{3\}}$) associated with the choice of the horizon values ($a_{x}^{I}$,$b_{x}^{I}$,$\Phi_x^{I},h_{tx}^{I}$) with the constraint.
Note that the upper index in the braces denotes the label of the solution basis, here.
We write these as the $3\times 4$ matrix
\begin{equation}
    H(r) :=
    \begin{pmatrix}
        \delta\Psi^{\{1\}} & \delta\Psi^{\{2\}} & \delta\Psi^{\{3\}}
    \end{pmatrix}.
\end{equation}

In the boundary constraint method, we also impose the constraint (\ref{eq:boundary_constraint}) on the boundary values.
To find the linear combination satisfying the constraint, it is convenient to consider the following combination
\begin{equation}
    F(\omega, r) :=
    m h_{tx}(r) + i \omega r_0^2 \Phi_x(r),
\end{equation}
which gives the left-hand side of (\ref{eq:boundary_constraint}) near the boundary.
We define the $3\times3$ matrix
\begin{equation}
    \tilde{H}(r) :=
    \begin{pmatrix}
        a_x^{\{1\}} & a_x^{\{2\}} & a_x^{\{3\}}\\
        b_x^{\{1\}} & b_x^{\{2\}} & b_x^{\{3\}}\\
        F^{\{1\}} & F^{\{2\}} & F^{\{3\}}
    \end{pmatrix}.
\end{equation}
Using this matrix, we obtain the coefficients of the linear combination by $\tilde{H}^{-1}(r_b)$ with a large cutoff $r_b$.
We can construct the $3\times 4$ solution matrix as
\begin{equation}
    \begin{pmatrix}
        \delta\Psi^{\{a_x\}}&
        \delta\Psi^{\{b_x\}}&
        \delta\Psi^{\{F\}}
    \end{pmatrix}
    := H(r)\hat{H}^{-1}(r_{b}).
\end{equation}
The solution $\delta\Psi^{\{a_x\}}$ satisfy the Dirichlet conditions $a_{x}^{(0)}=1$, $b_{x}^{(0)}=0$ and the constraint (\ref{eq:boundary_constraint}) with $m \hat{h}_{tx}^{(0)}=0$.
$\delta\Psi^{\{F\}}$ satisfies $a_{x}^{(0)} = b_{x}^{(0)}=0$ and $m \hat{h}_{tx}^{(0)}=1$.
For example, the electric conductivity $\sigma_A$ is obtained by using $\delta\Psi^{\{a_x\}}$.

In the RGS method, we consider the extra solution (\ref{eq:extra_solution}) as one of the solution basis.
Writing $\delta\Psi^{\{4\}} = (0, 0, m\zeta_0, -i\omega \frac{r_0^2}{r^2} g(r) \zeta_0)^{T}$, we can make the $4\times 4$ matrix
\begin{equation}
    H_{\text{RGS}}(r) :=
    \begin{pmatrix}
        \delta\Psi^{\{1\}} &
        \delta\Psi^{\{2\}} &
        \delta\Psi^{\{3\}} &
        \delta\Psi^{\{4\}}
    \end{pmatrix},
\end{equation}
where $\zeta_0$ is arbitrary nonzero constant.
Similar to the previous method, $H_{\text{RGS}}^{-1}(r_b)$ gives the coefficients of the linear combination.
Actually, taking the linear combination with $\delta\Psi^{\{4\}}$ implies considering the residual gauge transformation.
We obtain the $4\times 4$ solution matrix as
\begin{equation}
    \begin{pmatrix}
        \delta\Psi^{\{a_x\}}_{\text{RGS}} &
        \delta\Psi^{\{b_x\}}_{\text{RGS}} &
        \delta\Psi^{\{\Phi_{x}\}}_{\text{RGS}} &
        \delta\Psi^{\{h_{tx}\}}_{\text{RGS}}
    \end{pmatrix}
    :=
    H_{\text{RGS}}(r) H_{\text{RGS}}^{-1}(r_b).
\end{equation}
Each solution simply satisfy the diagonal Dirichlet boundary conditions at the boundary.
We can compute several conductivities by using these solutions.
For example, the electric conductivity can be obtained by using $\delta\Psi_{\text{RGS}}^{\{a_x\}}$.
As we have mentioned, the results of the conductivities are same as those obtained in the first method.

\section{Symmetry between conductivities in normal phase}
\par
\begin{figure}[htpb]
\begin{centering}
\subfigure[ Re($\sigma_A$) and Re($\sigma_B$) at the doping parameter x=1.]{
			\label{fig:Figure9a}
{\includegraphics[scale=0.75]{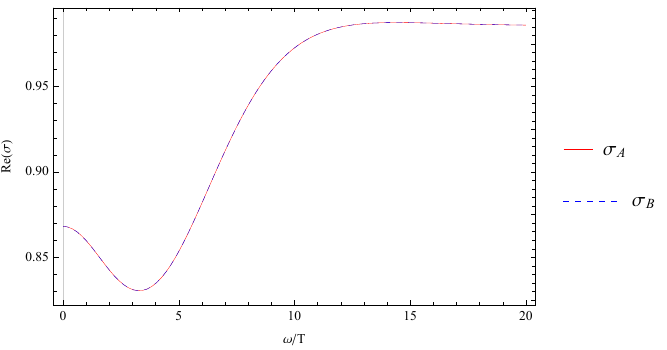}}}
\subfigure[ Re($\gamma$) at the doping parameter x = 0.5 and x=2.)]{
			\label{fig:Figure9b}
{\includegraphics[scale=0.75]{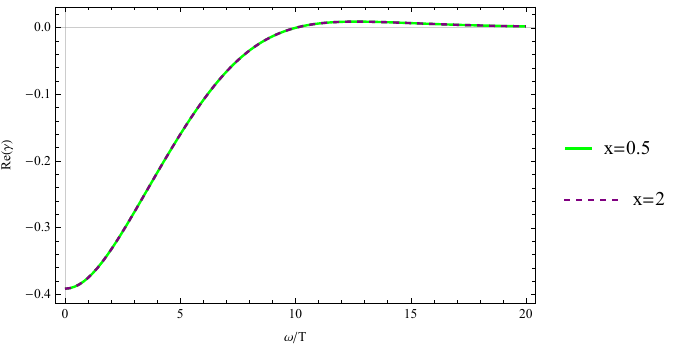}}}
\par\end{centering}
\caption{\label{fig:Figure9} A case without momentum dissipation $m=0$ at $T=0.33$, the two lines coincide perfectly. The system enjoys the symmetry. There is a delta function at $\omega=0$ but not shown.}
\end{figure}

\begin{figure}[htpb]
\begin{centering}
\subfigure[ The conductivity $\sigma_A$ and $\sigma_B$ at the doping parameter x=1.]{
			\label{fig:Figure5a}
{\includegraphics[scale=0.75]{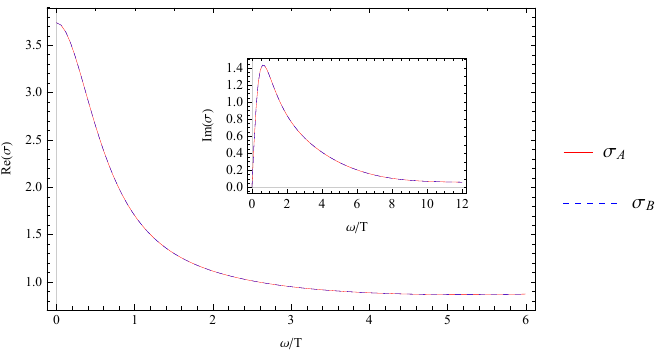}}}
\subfigure[ The conductivity $\gamma$ at the doping parameter x = 0.5 and x=2.)]{
			\label{fig:Figure5b}
{\includegraphics[scale=0.75]{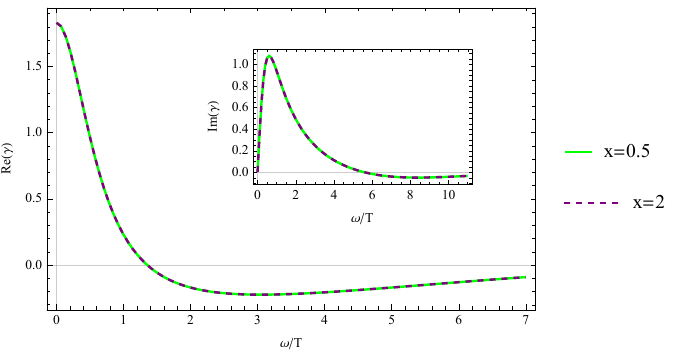}}}
\par\end{centering}
\caption{\label{fig:Figure5} A case with momentum dissipation $m=1$ at $T=0.33$, the two lines coincide perfectly. The system still enjoys the symmetry.}
\end{figure}

 In this section, We present electrical conductivity in our model that support the conclusions that the conductivities $\sigma_A$, $\sigma_B$ and $\gamma$ satisfy the symmetry \eqref{eq27}.
In the case of no momentum dissipation, we show the real part of the conductivity $\sigma_A$ and $\sigma_B$ at $\text{x}=1$ in Fig.\ref{fig:Figure9a}, $\gamma$ at x=0.5 and x=2 in Fig.\ref{fig:Figure9b}, respectively. Figure \ref{fig:Figure5} shows images of the conductivities with the momentum dissipation parameter $m=1$. The fact that two lines coincide perfectly prove the symmetry of the system still exists after the translation symmetry is broken.

\bibliographystyle{unsrt}
\bibliography{ref}

\end{document}